\documentclass[a4paper]{jpconf}
\usepackage{graphicx}
\usepackage{subcaption}
\usepackage[hidelinks,
colorlinks = true,
urlcolor=blue,
linkcolor = blue,
citecolor = blue]{hyperref}
 
\begin{document}
\title{Standalone, Descriptive, and Predictive Digital Twin of an Onshore Wind Farm in Complex Terrain}
\author{{Florian Stadtmann}$^1$, {Adil Rasheed}$^{1,2}$, {Tore Rasmussen}$^3$}
\address{$^1$ Department of Engineering Cybernetics, Norwegian University of Science and Technology, Trondheim, Norway}
\address{$^2$ Department of Mathematics and Cybernetics, SINTEF Digital, Trondheim, Norway}
\address{$^3$ Aneo, Norway} 

\ead{florian.stadtmann@ntnu.no}

\begin{abstract}
In this work, a digital twin with standalone, descriptive, and predictive capabilities is created for an existing onshore wind farm located in complex terrain. A standalone digital twin is implemented with a virtual-reality-enabled 3D interface using openly available data on the turbines and their environment. Real SCADA data from the wind farm are used to elevate the digital twin to the descriptive level. The data are complemented with weather forecasts from a microscale model nested into Scandinavian meteorological forecasts, and wind resources are visualized inside the human-machine interface. Finally, the weather data are used to infer predictions on the hourly power production of each turbine and the whole wind farm with a 61~hours forecasting horizon. The digital twin provides a data platform and interface for power predictions with a visual explanation of the prediction, and it serves as a basis for future work on digital twins.
\end{abstract}

\section{Introduction}
    The importance of wind energy production efficiency cannot be overstated in the context of combating climate change and achieving a net-zero emissions target by 2050 \cite{2018ida}. With the proliferation of cheaper sensors and the growing trend of the Internet of Things, the potential for extracting data from wind farms has increased significantly. However, it is not sufficient to store collected information in data silos. Instead, real-time data analysis and visualization can be leveraged to enable optimal control and informed decision-making and to unlock the full potential of the data.

    The concept of the digital twin has emerged as a promising solution to address these challenges. A digital twin utilizes available data in real-time to monitor the current state of an asset and its environment, predict future states, detect faults, perform what-if scenario analysis, provide decision support, and ultimately enable autonomous control of the asset~\cite{Rasheed2020dtv}. The use of a suitable human-machine interface enhances the interpretation of analysis results and allows for effective communication with stakeholders.

    A survey conducted with industry partners of the Norwegian Research Centre on Wind Energy “FME NorthWind” indicates that the wind industry is keenly interested in utilizing digital twins to reduce the cost of wind energy~\cite{Stadtmann2023dti}. However, several challenges must be addressed before the full potential can be unlocked in wind energy applications. These challenges relate to both the implementation and acceptance of digital twins within the industry~\cite{Stadtmann2023dti}. Overcoming these challenges will be critical to advancing the development and adoption of digital twins in the wind energy sector.

    To this end the current work attempts to realize the following:
    \begin{itemize}
        \item Introducing readers to the concept of digital twins within the context of wind energy applications and providing a scale to rank digital twins based on their capabilities.
        \item Demonstrating a digital twin of an onshore wind farm with standalone, descriptive, and predictive capabilities. This will provide a practical illustration of the potential benefits of digital twins in wind energy applications, as well as offer insights into the challenges of developing such models.
        \item Discussing the potential for further research on digital twins for wind farm applications. By highlighting areas where additional research is needed, we hope to catalyze progress in this field and drive innovation in the wind energy sector.
    \end{itemize} 
    
    The article is structured as follows:
    First, the definition of the term digital twin used in this work is clarified in section~\ref{sec:concept}. The capability level scales are explained briefly. In section~\ref{sec:standalone}, the implementation of the standalone digital twin is given with a focus on terrain and visual interface. The onshore Bessakerfjellet wind farm is used as a demonstration site. It is operated by Aneo and is located at (64°13'~N,~10°23'~E) on the Norwegian coastline. Section~\ref{sec:descriptive} explains the integration and visualization of data measured at the turbines. Predictive capabilities are added in section~\ref{sec:predictive} by implementing weather forecasts and performing predictions of the wind turbines' power production.
    The work is discussed in section~\ref{sec:discussion} and an outlook into future work is given. Finally, the work is summarized in section~\ref{sec:conclusion}.
    \begin{figure}[h!]
            \centering
            \begin{minipage}[t]{0.8\textwidth}
            \centering
                \includegraphics[width=\textwidth]{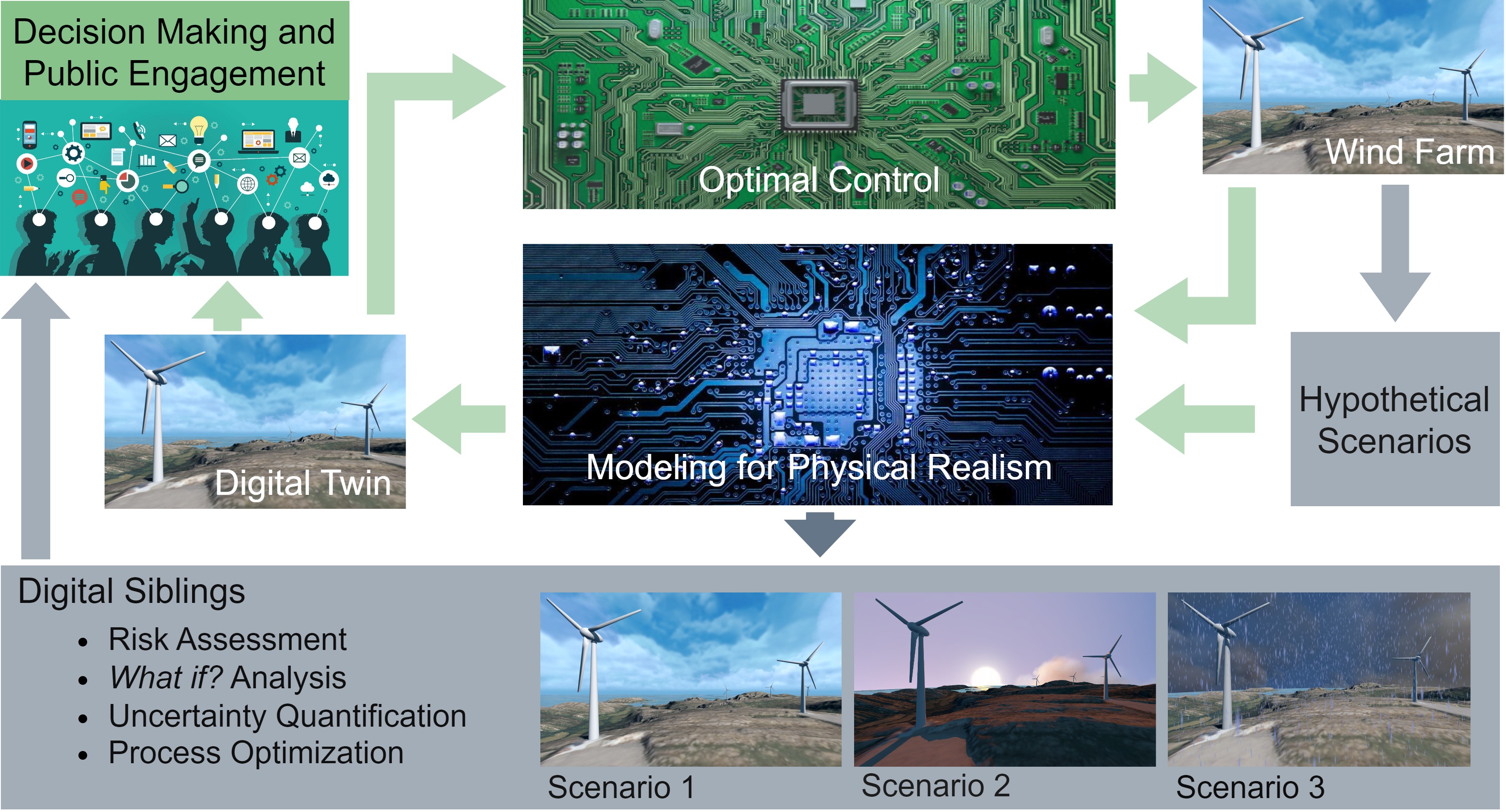}
            \captionof{figure}{Digital twin flowchart (adapted from \cite{Pawar2021haa})}
            \label{fig:dt}
            \end{minipage}
            \hspace{0.01\textwidth}
    \end{figure}

\section{Definition and Capability Levels}
\label{sec:concept}
\begin{figure}[h!]
            \centering
            \begin{minipage}[t]{0.8\textwidth}
            \centering
                \includegraphics[width=\textwidth]{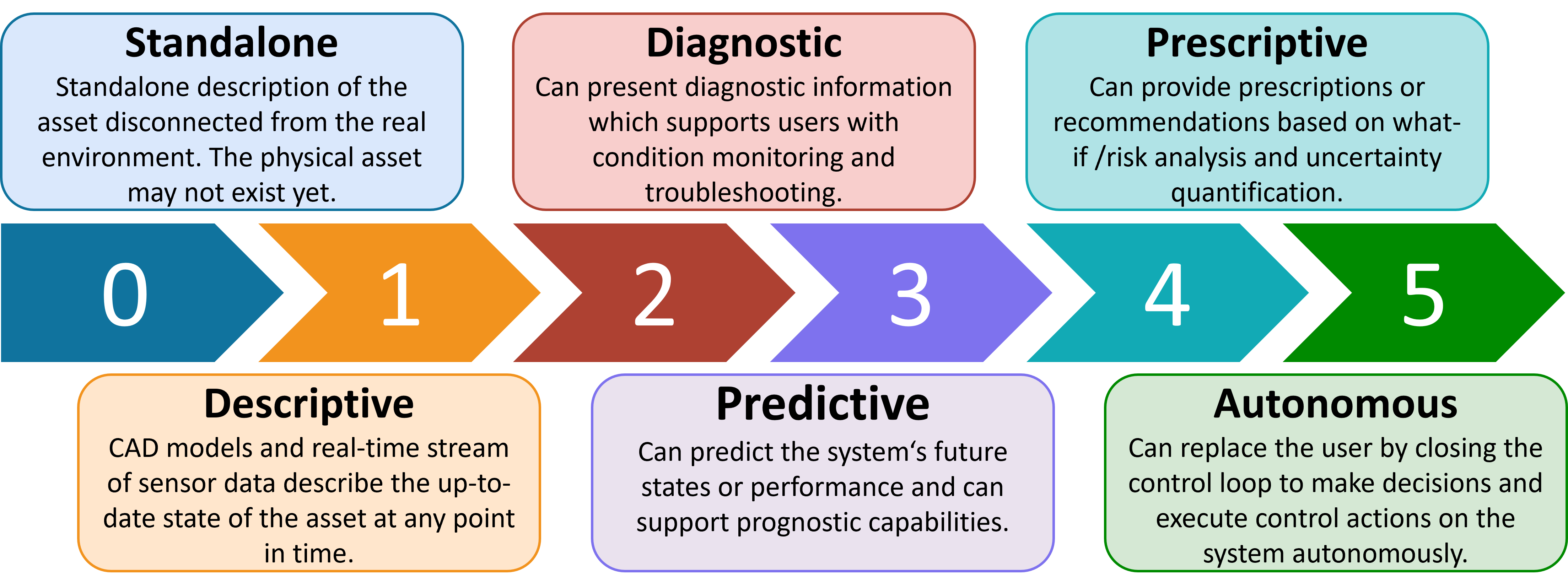}
            \captionof{figure}{Capability level scale (adapted from \cite{San2021haa})}
            \label{fig:cap}
            \end{minipage}
    \end{figure}

    The term digital twin is being used for different concepts. Here, the digital twin is "a virtual representation of a physical asset or a process enabled through data and simulators for real-time prediction, optimization, monitoring, control, and informed decision making"~\cite{Rasheed2020dtv}. 
    The concept of a digital twin with all capabilities is shown in figure~\ref{fig:dt}.
    Since this definition still leaves some room, we use the capability level scale from~\cite{San2021haa} to specify the digital twin's exact capabilities. As such, a digital twin can be ranked on a scale from 0 to 5 as a standalone, descriptive, diagnostic, predictive, prescriptive, or autonomous digital twin as shown in figure~\ref{fig:cap}. Here, a brief overview of the capability level scale is given. A more detailed description can be found in~\cite{Stadtmann2023doa} and~\cite{Stadtmann2023dti}.
    \begin{itemize}
        \item The \textbf{standalone} digital twin is a virtual representation of a wind farm that lacks a real-time connection to the physical wind farm. It can be utilized in the design, planning, and construction stages before the wind farm is operational.
        \item In the \textbf{descriptive} digital twin, measurements from the wind farm are being streamed into the digital twin. The descriptive digital twin mirrors the state of the real wind farm at each point in time and provides a platform on which data can be bundled, enhanced (e.g. through virtual sensing), processed, and visualized to the human operators and other stakeholders.
        \item The \textbf{diagnostic} digital twin uses the data gathered in the descriptive digital twin as input for analysis such as condition-based maintenance. The condition of components is tracked through e.g. vibration and temperature measurements, and anomalies are diagnosed. This way, minor deficits can be detected early and resolved before they result in major faults like turbine damage and unexpected downtime.
        \item A \textbf{predictive} digital twin does not only use current and historical data but also forecasts parameters to predict future asset states. The predictive capabilities can be used for predictive maintenance or through power forecasts for the energy market.
        \item In a \textbf{prescriptive} digital twin, recommendations are provided through what-if scenario analysis and risk assessment. Such prescriptions can include a balancing of component wear against power production based on current electricity prices and demand, or optimal maintenance scheduling based on component wear, estimated remaining useful lifetime, data anomalies, and weather forecasts. 
        \item The \textbf{autonomous} digital twin acts on the prescriptions on its own. Autonomous digital twin capabilities can range from farm-wide wake steering and component wear balancing over inspection through the usage of autonomous drones to automated operation and maintenance of the wind farm.
    \end{itemize}

\section{Standalone Digital Twin}
    \label{sec:standalone}
    In this work, a standalone digital twin of the onshore wind farm has been implemented following a similar approach as is explained in~\cite{Stadtmann2023doa} for a floating offshore wind turbine, including a user interface using virtual reality. In contrast to the single-turbine implementation in~\cite{Stadtmann2023doa}, a whole wind farm is implemented here. Additionally, the local terrain around the wind farm is included. 
    
    \subsection{Terrain}
        As an input to the terrain generation, height maps of the local terrain are being downloaded from~\cite{hoydedata} at a $1~\mathrm{m} \times 1~\mathrm{m}$ resolution grid. Since the wind farm is located at a shoreline and the LIDAR-based height measurements cannot penetrate the water surface, the height maps are being complemented with information on ocean depth contour lines available from~\cite{kartverket}. All information on onshore and offshore terrain height is combined in a single terrain map. The height is then binned into int16 and the map is split into equal chunks to improve computational efficiency during rendering. Next, aerial images are downloaded from~\cite{norgeibilder} with a $1~m \times 1~m$ resolution. The images are combined and split into chunks matching the chunks of the terrain height. Terrain height and texture are then imported into the Unity game engine, where they are combined. As evident from figure~\ref{fig:terrain}, a top-down view of the 3D terrain inside the game engine (center) can only be distinguished from aerial images from~\cite{norgeibilder}(surrounding) by its improved resolution, 3D terrain, animated water, and dynamic lighting. Note that the terrain is not just implemented for visual realism while using the digital twin, it also contains information on logistical access through roads and nearby villages, and information on terrain height, water bodies, and forestation relevant for understanding wind flow.
        
    \subsection{Turbines}
\begin{figure}[h!]
            \centering
            \begin{minipage}[t]{0.48\textwidth}
            \centering
                \includegraphics[width=\textwidth]{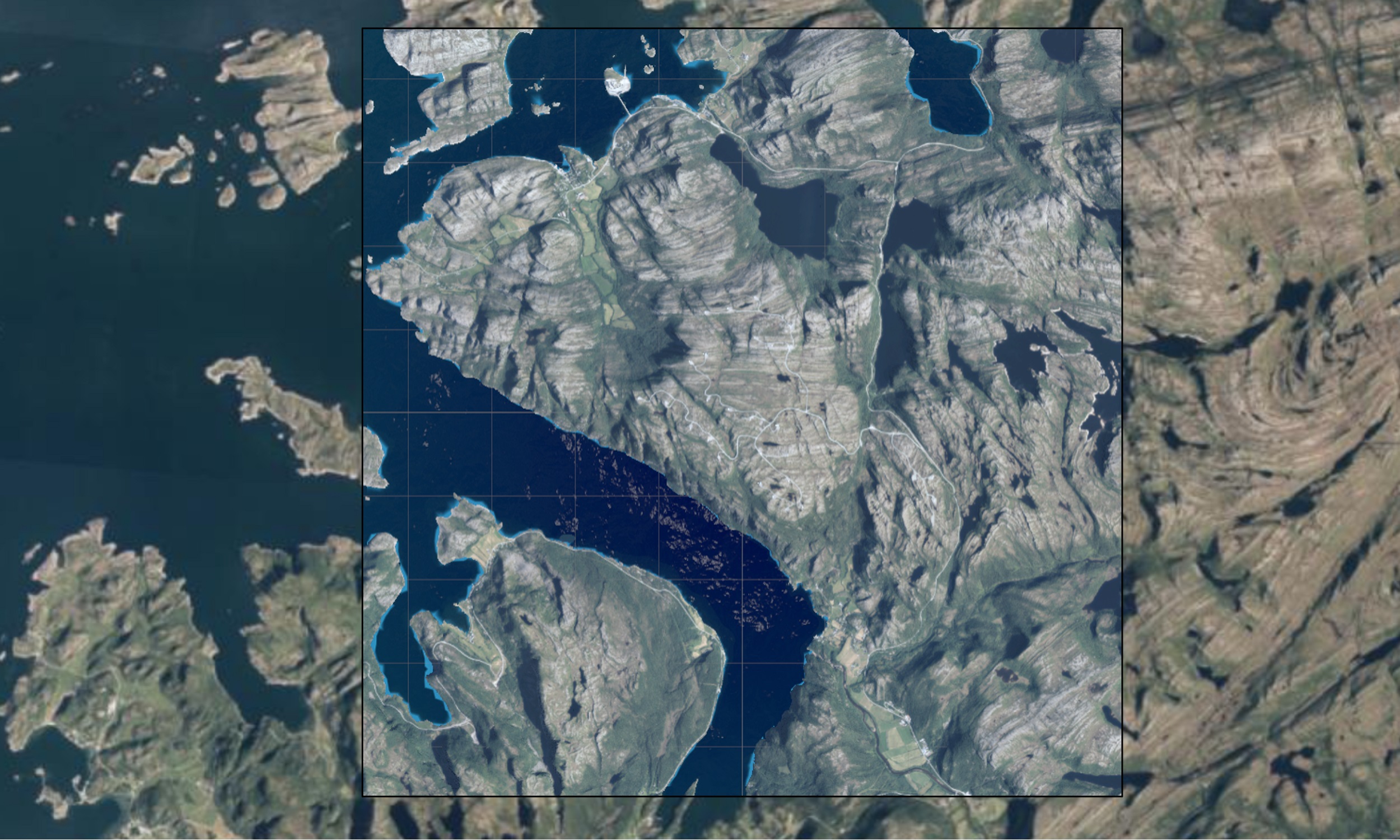}
            \captionof{figure}{Comparison of terrain between top-down view within the digital twin interface (inside) to a picture from~\cite{norgeibilder} (outside)}
            \label{fig:terrain}
            \end{minipage}
            \hspace{0.01\textwidth}
            \begin{minipage}[t]{0.48\textwidth}
            \centering
                 \includegraphics[width=\textwidth]{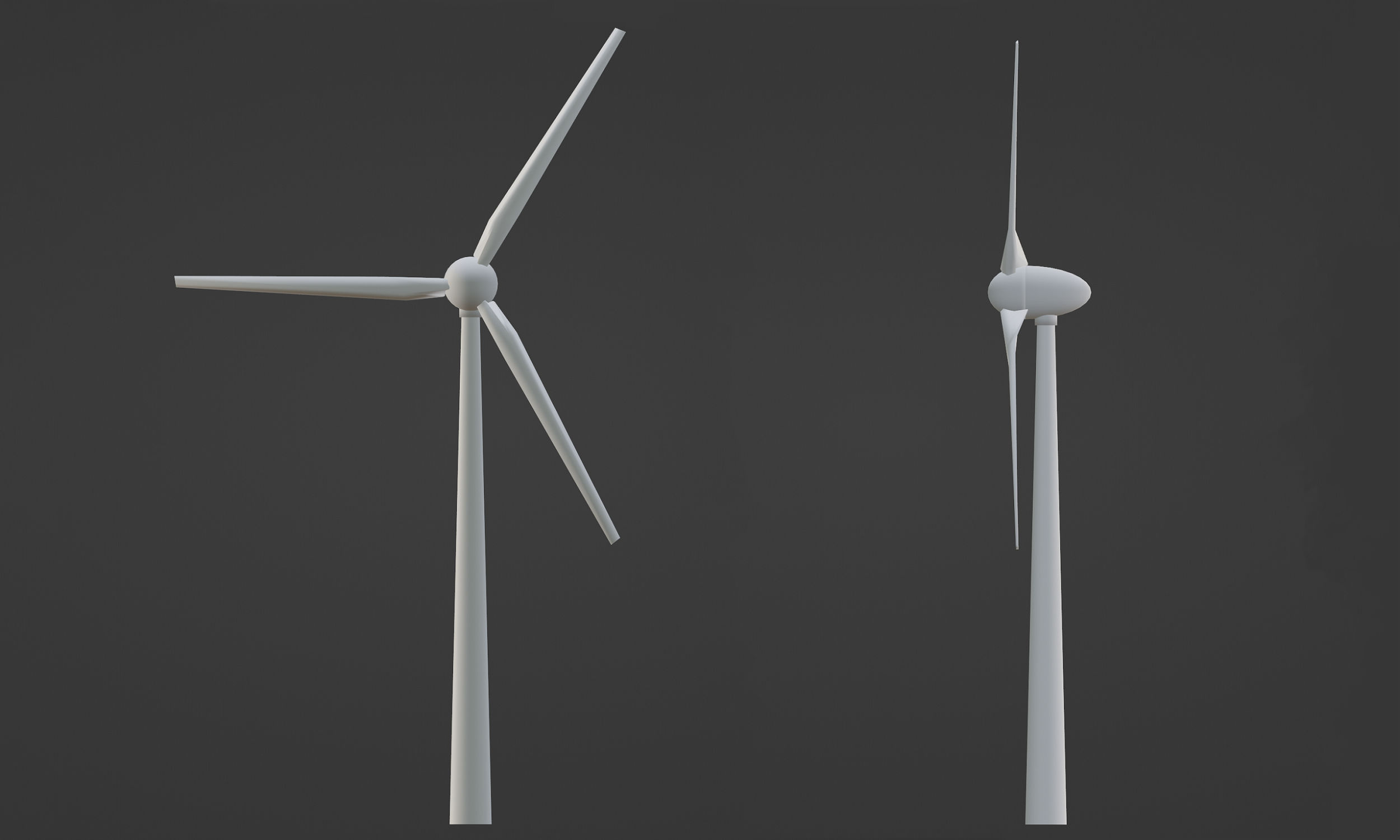}
            \captionof{figure}{3D CAD Model of the \\Enercon E70-4 turbine}
            \label{fig:turbine}
            \end{minipage}
        \end{figure}

        Since no CAD model of the turbines was available at the start of the project, a model was created in Blender. Tower height and rotor diameter are based on data sheets, while the nacelle and blades are based on pictures of the {Enercon E70-4}. The 3D CAD model of the turbine is shown in figure~\ref{fig:turbine}.
        The horizontal position of each turbine is known, while the vertical position is inferred from the terrain height.

\section{Descriptive Digital Twin}
    \label{sec:descriptive}
    The digital twin is enhanced with descriptive capabilities by including SCADA data from each wind turbine. At this stage, the digital twin mirrors the state of the physical wind turbine. Only minor changes were made to the implementation in~\cite{Stadtmann2023doa}. Namely, the data structure gained an additional hierarchical level to advance from turbine to farm-level and the interactable components of the turbine were adjusted to the new turbine type. Additionally, the data input format changed, which required rebuilding the data reading module. Finally, two visualization methods were added to depict the current power production, as it cannot be directly seen on the turbine models.
    
    \subsection{Data}
        The available data consist of wind speed, wind direction, nacelle direction, and active power from each turbine. The measurement intervals vary from 3 to 10~minutes. Since the data are non-equidistant in time, an updating function is constantly checking for new measurements. For real-time operation, the persistence method is used to bridge time spans without measurements. If instead the digital twin is used to inspect historic data, they are interpolated between measurements. The feasibility of real-time data streaming is demonstrated as explained in section~\ref{sec:predictive}.
        
    \subsection{Visualization}
\begin{figure}[h!]
            \centering
            \begin{minipage}[t]{0.8\textwidth}
                \includegraphics[width=\textwidth]{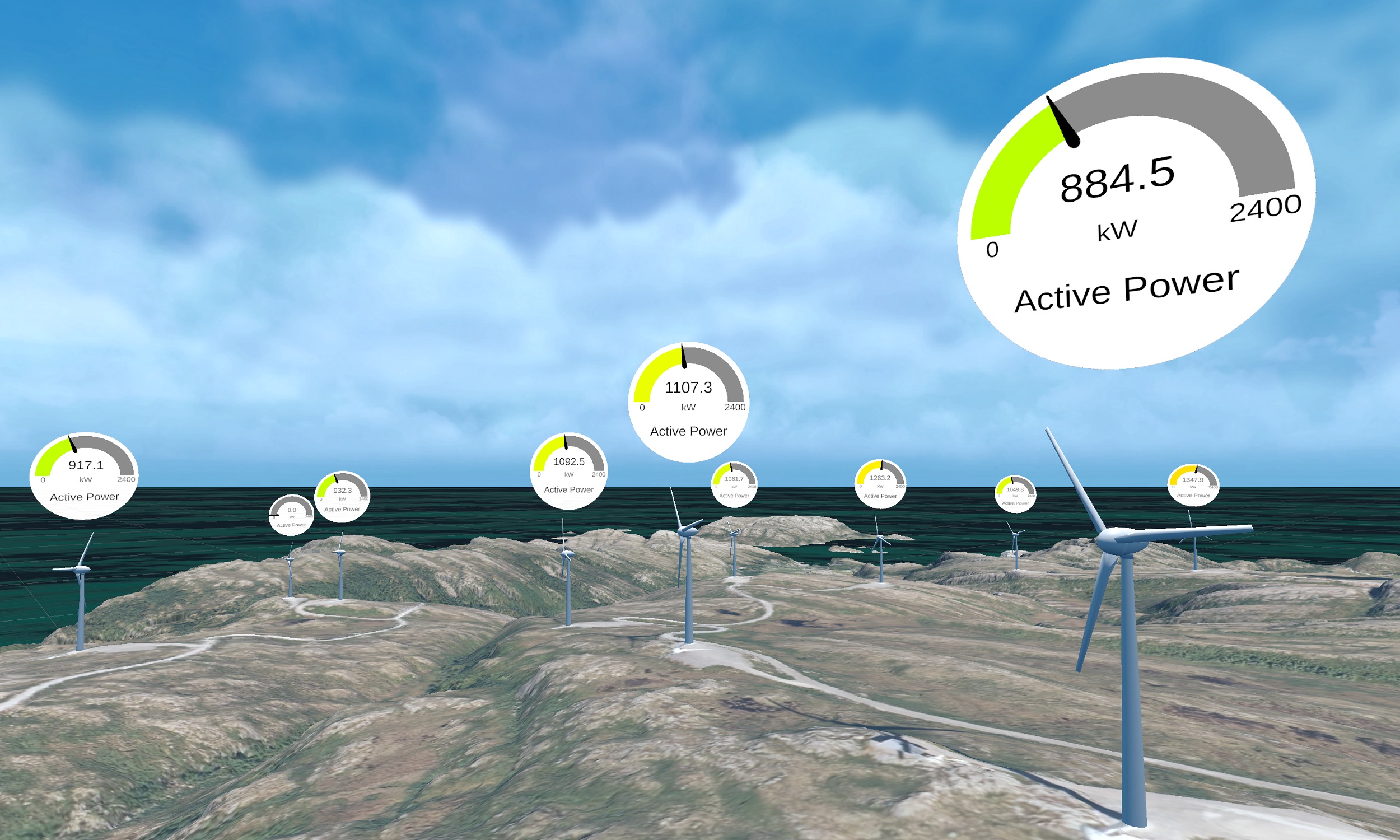}
                \captionof{figure}{Descriptive digital twin with text and gauges showing active power of each turbine.}
                \label{fig:gauges}
            \end{minipage}
        \end{figure}

        The yaw angle of each turbine is directly visible from the orientation of the turbine model. The active power can be shown in text above each turbine, or alternatively through gauges with dial and color indications as shown in figure~\ref{fig:gauges}.

\section{Predictive Digital Twin}
    \label{sec:predictive}
    First predictive capabilities are added to the digital twin by streaming publicly available weather forecast data. These external forecasts are then used to predict the theoretical power production at the turbine- and farm-levels.
    
    \subsection{Wind Field}
\begin{figure}[h!]
            \centering
            \begin{minipage}[t]{0.8\textwidth}
                \includegraphics[width=\textwidth]{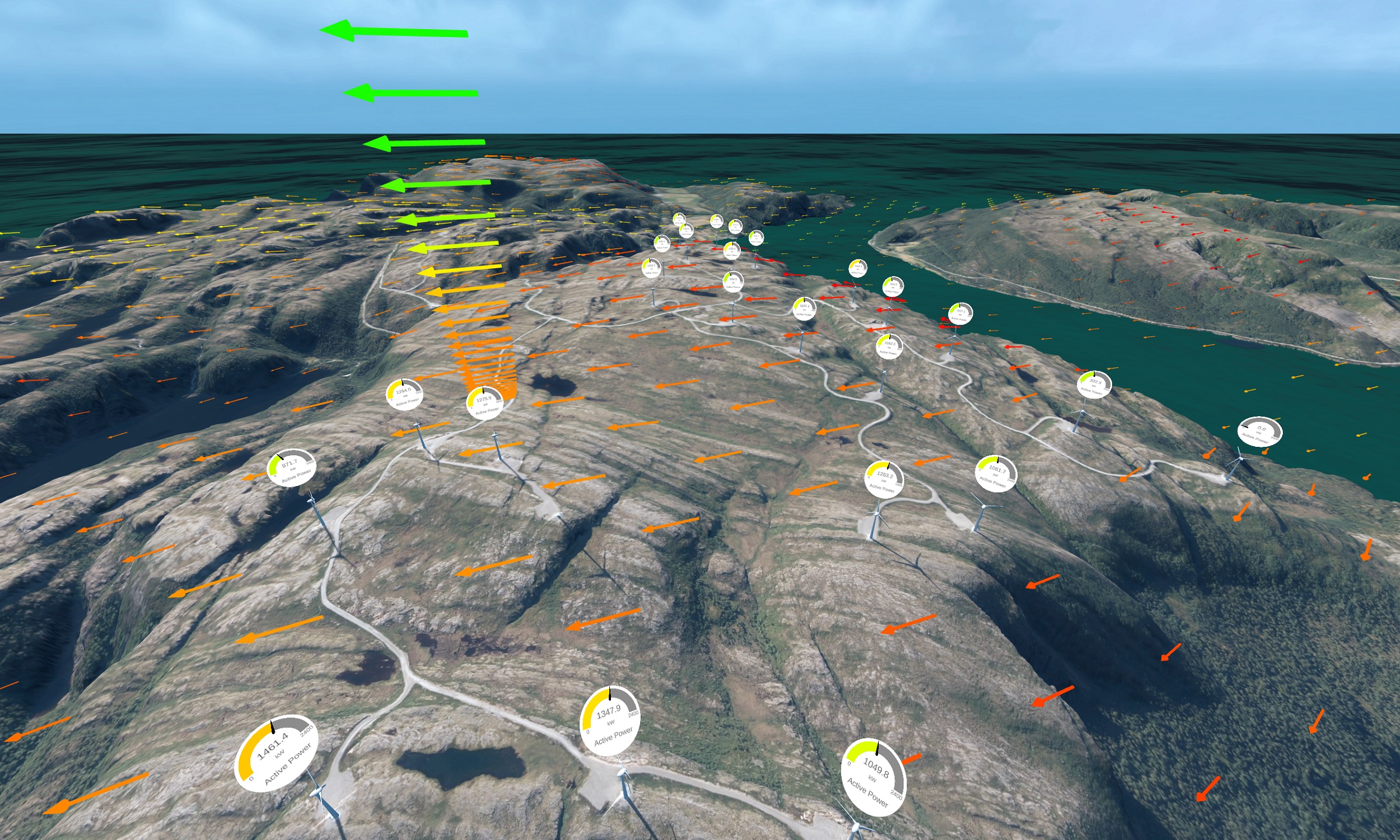}
                \captionof{figure}{Vector field and profile of wind with wind speed (length), wind direction (orientation), and turbulence (color) for any forecast horizon.}
                \label{fig:windfield}
            \end{minipage}
        \end{figure}

        A vector field for wind speed and direction is implemented by streaming weather forecasts from the Norwegian Meteorological Institute's Thredds service \cite{metthredds} in real-time. The MetCoOp Ensemble Prediction System (MEPS) \cite{NorwegianMeteorologicalInstitutemd} provides forecasts every 6~h up to 61~h ahead with a frequency of 1~h. Parameters of interest are wind speed, wind direction, air pressure, air temperature, and relative air humidity. However, the MEPS model has a resolution of only 2.5~km. For this reason, the SIMRA microscale model nested into the HARMONIE mesoscale model is used around the wind farm to increase the lateral and vertical resolution of the forecast and include effects induced through the complex terrain. More information on the HARMONIE and SIMRA models can be found in~\cite{Rasheed2017wfm}. The SIMRA model evaluated around the Bessakerfjellet wind farm is available at \cite{metthredds}. It includes wind speed, wind direction, air pressure, and air temperature in 1~h intervals for 6~h to 18~h ahead and has been evaluated every 12~h for this particular data set. There is no technical reason preventing the SIMRA model from being evaluated more frequently and with a longer forecasting horizon apart from saving on computational resources. The wind is visualized in the digital twin through wind trails moving through the vector field, or by showing parts of the vector field directly as can be seen in figure~\ref{fig:windfield}. Vector direction matches wind direction, vector length represents wind speed, and color indicates the turbulence index.
        
    \subsection{Physics-based power prediction}
        In the next step, the weather forecast is used to estimate the power production at each turbine. In~\cite{Yakoub2023ieo}, weather forecasts such as the MEPS forecast were used with support vector machines, clustering methods, and random forest algorithms to map from wind to power production in flat terrain. Here, the weather forecast is used as input, but the mapping from weather to produced power is done through physics-based models (PBMs) only to circumvent the black-box problem of data-driven methods (DDMs). 
        A data sheet for the turbine type is used that contains the direct mapping from wind speed to power production, as well as the power coefficient as a function of wind speed, with $1~\frac{\mathrm{m}}{\mathrm{s}}$ intervals. The power coefficient can be used in the well-known relation 
        \begin{equation}
        \label{eq:power_curve}
            P(v)=\frac{1}{2} \rho C_P(v) A v^3
        \end{equation}
        where $v$ is the wind speed, $P(v)$ is the produced power, $\rho$ is the density of the air, $C_P(v)$ is the turbine-specific power coefficient, and $A$ is the area swept by the blades.
        The blade sweeping area $A$ is known to be $3959~\mathrm{m}^2$.
        In the first approach, the air density $\rho$ is assumed to be constant with $\rho_s=1.225~\frac{\mathrm{kg}}{\mathrm{m}^3}$. However, air density depends on temperature and pressure. Treating air as an ideal gas, the pressure of dry air $\rho(T,p)$ can be calculated as
        \begin{equation}
            \rho(T,p) = \frac{p M_{d}}{RT} 
        \end{equation}
        where $T$ is the air temperature, $p$ is the pressure of the air, $M_d=0.0289652~\frac{\mathrm{kg}}{\mathrm{mol}}$ is the molar mass of dry air, and $R=8.31446~\frac{\mathrm{J}}{\mathrm{K}~\mathrm{mol}}$ is the universal gas constant. Furthermore, humidity can be included through 
        \begin{equation}
        \rho(T,p,\phi) = \frac{(p- \phi p_{sat}) M_d + \phi p_{sat} M_v}{RT} 
        \end{equation}
        with $M_v=0.018016~\frac{\mathrm{kg}}{\mathrm{mol}}$ as the molar mass of water vapour, $\phi$ as the relative humidity, and 
        \begin{equation}
            p_{sat} = 6.11~hPa *10^\frac{7.5 (T+273.15~\mathrm{K})}{T+510.45~\mathrm{K}}
        \end{equation}
        as the saturation pressure, calculated with the Tetens equation as done in~\cite{Bolton1980tco}.
        The air density can be used to modify the power curve by
        \begin{equation}
            P(v,T,p,\phi) = P(v) \frac{\rho(T,p,\phi)}{\rho_s}
        \end{equation}
        
        The quality of the power prediction is calculated on a one-year training set with an hourly resolution for each combination of 
        \begin{itemize}
            \item wind speed, air temperature, and air pressure $v$ based on the MEPS or SIMRA model,
            \item air density $\rho$ constant, dry air, or humid air (for SIMRA-based models only),
            \item calculation from the power curve or through the power coefficient from the turbine manufacturer and equation~\ref{eq:power_curve},
            \item interpolation of power curve or power coefficient with linear or cubic,
            \item with or without imposing an upper limit on power output according to turbine specifications.
        \end{itemize}

    \subsection{Data-Driven Predictions}
        Purely data-driven predictions using dense neural networks (DNN) and long-short-term-memory (LSTM) neural networks are implemented for measurement-based time series prediction and compared with the results from the PBMs.
        In the DDMs, two years of data are being used to train the neural networks (NNs), where 10\% are split off for validation. The NNs are being trained for one-step-ahead prediction of the power production, and are evaluated iteratively on their output to obtain a forecast with the full 61~h forecasting horizon. Therefore, the NN output is of size 1. The architecture of the NNs is kept simple with three layers with 5, 3, and 1~units respectively. The input lag is chosen to be 4~h for the DNN based on the partial autocorrelation. In contrast, the cells of the LSTM keep information from previous evaluations in memory. Therefore, only one input is given at a time but the NN is evaluated on a sequence of previous data points. The NNs are trained with the Adam optimizer with a default learning rate, a batch size of 64, and the mean squared error as the loss metric. A validation-loss-based early stopping is used to avoid overfitting. 
        Since the partial autocorrelation suggests that the last measured hour has by far the most substantial contribution to the short-term prediction, the NNs are compared to the persistence model, which always predicts the last measured value.
        
    \subsection{Results}
\begin{figure}[h!]
            \centering
            \begin{minipage}[t]{0.48\textwidth}
            \centering
                 \includegraphics[width=\textwidth]{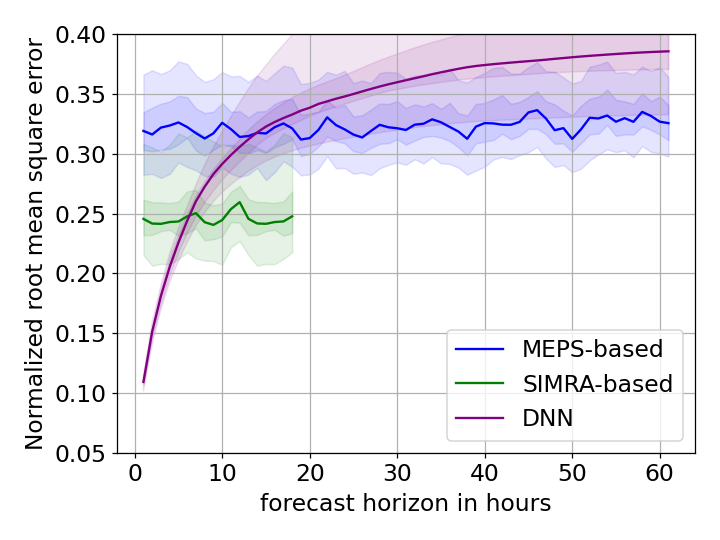}
            \captionof{figure}{Turbine level NRMSE of the best model in each category across $61~h$ forecasting horizon. Solid: median turbine, transparent: 1. \& 3. quantile and 
            extrema.
            }
            \label{fig:turbines}
            \end{minipage}
            \hspace{0.01\textwidth}
            \begin{minipage}[t]{0.48\textwidth}
            \centering
                \includegraphics[width=\textwidth]{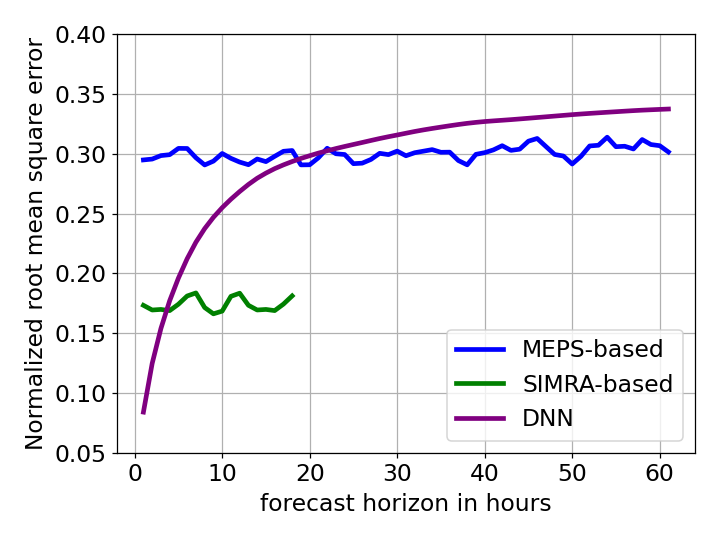}
            \captionof{figure}{Farm-level NRMSE of the best model in each category}
            \label{fig:farm}
            \end{minipage}
        \end{figure}

        The PBMs and DDMs are compared against the measured power production for every single turbine and for the whole farm production by using the normalized root mean squared error (NRMSE) for 3~years of available data. The best-performing model in each category is determined. The NRMSE across turbines is shown in figure~\ref{fig:turbines}, and the NRMSE on farm-level prediction in figure~\ref{fig:farm}. Note that the farm-level predictions are more accurate as prediction errors of different turbines can cancel each other. The DDMs perform best for small forecast horizons, but their accuracy decays quickly. For one-hour-ahead forecasts, the DNN performs marginally better than the LSTM and persistence model with $<$0.2\%~NRMSE. 
        The SIMRA models outperform the DNN after 2~h on the farm-level and after 5~h on the turbine level. All predictions using the SIMRA model as input achieve similar accuracy, but a dynamic air density does improve the forecast by 0.4\%~NRMSE. The improvement from dry to humid air density and differences between the interpolation method, the cap on maximum power production, and the difference between using the calculated power curve or power coefficient as input are much smaller with $<$0.1\%~NRMSE. 
        The best-performing SIMRA-based model uses the turbine's power curve with cubic interpolation, a limit on the maximum power production as the rated power, and a correction for air density that accounts for humid air.
        The micro-scale SIMRA model gives significantly better results than the MEPS model. Here the SIMRA model was only available up to $18~h$ ahead, but it is expected that the SIMRA-based models will keep outperforming the MEPS-based models also for longer forecasting horizons as the decay of accuracy with increasing prediction horizon is slow.
        Differences within the MEPS models are small $<$0.12\%~NRMSE. Like the SIMRA-based models, the best-performing MEPS-based model uses the power curve directly with cubic interpolation.
    
        The different models can be combined in a simple hybrid analysis and modeling (HAM) approach for optimal wind farm power prediction on all forecasting horizons by using the DNN for 1~h to 2~h ahead predictions, the SIMRA-based model for 3~h to 18~h ahead predictions, and the MEPs-based model for 19~h to 61~h ahead predictions.
        Deriving the farm-level power forecast from the turbine level forecasts makes it possible to assess the impact of each turbine on the farm power production separately inside the virtual-reality-enabled interface and visually trace reasons for fluctuations between turbines back to wind speed, direction, and turbulence, as well as to terrain geometry and surface roughness. Therefore, the digital twin can be used to explain the farm-level power forecast.

\section{Discussion and Future Work}
    \label{sec:discussion}
    In this work, a functional digital twin of a wind farm was presented with standalone, descriptive, and predictive capabilities. 
    There is much room for further research and demonstration on all capability levels. Hence, this section discusses potential improvements and future work.  
    
    \subsection{Standalone}
        The modular nature of the implementation allowed upscaling the standalone digital twin from the turbine level presented in~\cite{Stadtmann2023doa} to the farm-level without any difficulties, but a terrain had to be added for the onshore wind farm. In this work, the aerial images were directly mapped onto the terrain, resulting in 81~big textures with 1~m resolution and ca. 1~$\mathrm{km}^2$ area. This did not compromise the real-time execution of the digital twin, but it may be advantageous to use one smaller, high-resolution texture per land cover class (grass/moss/rock/forest/etc.) instead. In future work, their placement can be automated by compiling the color channels of the aerial images into arrays with texture information. The software could be extended to include automated placing of forestation and houses. This way, the detail in the visual component of the digital twin can be increased, and the project size can be reduced. 
    
    \subsection{Descriptive}
        The upgrade to the farm-level required an additional level in the data hierarchy. A new visualization method was included in the digital twin to ease the assessment of data such as power production across the whole wind park. 
        The different data formats required a new interface between the raw data and the digital twin. Standardization will play an important role in the commercialization of digital twins, and more efforts are needed to establish common standards throughout the whole wind energy industry. Both wind turbine operators and original equipment manufacturers are already collaborating on standardization efforts \cite{ea,2022ain}
    
    \subsection{Predictive}
        For predictions in this work, three models have been combined in a pipeline where the best model is used depending on the time to be predicted ahead.
        More sophisticated HAM approaches could potentially improve the predictions further by combining information from measured data and numerical weather models. In the broader context of digital twins for wind energy, PBMs, DDMs, and HAMs have been discussed in \cite{Stadtmann2023dti}.
        In the context of wind power predictions, an example of a HAM approach includes a data-driven regression from mesoscale weather forecasts to power production, as has been evaluated for flat and open terrain in \cite{Yakoub2023ieo}. The microscale weather model could be replaced by a resolution-enhancing generative adversarial network, as has been attempted in \cite{Tran2020ges}, but the results were criticized in \cite{Larsen2020ota}.
        Finally, ensemble methods with secondary models for combining DDM and PBM outputs may extract additional value from measurements and numerical models. This approach will be investigated in future work with a more thorough investigation of different DDMs.

    \subsection{Diagnostic, Prescriptive, and Autonomous}
        In addition to the standalone, descriptive, and predictive capabilities explored here, diagnostic modules could be evaluated on the measured data for condition monitoring and component wear tracking including weather effects and turbine load. Prescriptive modules could include weather- and power-aware maintenance scheduling. Finally, the turbine state could be used for autonomous farm optimization and control to balance power production and turbine wear \cite{Kolle2022tiw}.
        
\section{Conclusion}
\label{sec:conclusion}
In this work, a digital twin of an onshore wind farm in complex terrain with standalone, descriptive, and predictive capabilities was built.
The standalone digital twin was implemented into a game engine by creating a 3D CAD model of the turbines and combining it with height maps and aerial images of the surrounding terrain. It includes a human-machine interface capable of interaction through virtual reality and simultaneously contains meta-data about the wind turbines, the farm layout, and the environment. The digital twin was elevated to the descriptive level by including SCADA data measured at each of the wind turbines. Some data were visualized directly on the 3D CAD model, while other information was shown with animated gauges and text. Predictive capabilities were implemented to forecast the power production of each turbine in the wind farm and the results were visualized in the interface. The predictions were performed by combining existing weather forecasts and physics-based models. They were made intuitively understandable by showing wind as vector fields and trails on the terrain. Finally, the results were discussed, and an outlook on future work was given. On top of the continuation of current research, the digital twin can be extended with additional modules to cover more aspects and evolve throughout the whole life-cycle of a wind farm.
Such full-fledged digital twins will have the potential to substantially contribute towards cheaper and more sustainable wind energy for a greener future.

\ack
This publication has been prepared as part of NorthWind (Norwegian Research Centre on Wind Energy) co-financed by the Research Council of Norway (project code 321954), industry, and research partners. Read more at \cite{FMENorthWindnrc}.
Data for the project were provided by Aneo. Mock data were used for the creation of figure~\ref{fig:gauges} and figure~\ref{fig:windfield} to comply with confidentiality agreements.

\section*{References}

\bibliographystyle{AR}
\bibliography{references_1}

\begin{thebibliography}{10}
\expandafter\ifx\csname url\endcsname\relax
  \def\url#1{\texttt{#1}}\fi
\expandafter\ifx\csname urlprefix\endcsname\relax\def\urlprefix{URL }\fi
\expandafter\ifx\csname href\endcsname\relax
  \def\href#1#2{#2} \def\path#1{#1}\fi

\bibitem{2018ida}
E.~Commission,
  \href{https://knowledge4policy.ec.europa.eu/publication/depth-analysis-support-com2018-773-clean-planet-all-european-strategic-long-term-vision_en}{In-depth
  {Analysis} in {Support} of the {Commission} {Communication} {COM}(2018) 773},
  Tech. rep., European Commission, Brussles (2018).

\bibitem{Rasheed2020dtv}
A.~Rasheed, O.~San, T.~Kvamsdal,
  \href{https://ieeexplore.ieee.org/document/8972429/}{Digital {Twin}:
  {Values}, {Challenges} and {Enablers} {From} a {Modeling} {Perspective}},
  {\em IEEE Access}, 8:21980--22012 (2020).
\newblock

\bibitem{Stadtmann2023dti}
F.~Stadtmann, A.~Rasheed, T.~Kvamsdal, K.~A. Johannessen, O.~San, K.~Kölle,
  J.~O.~G. Tande, I.~Barstad, A.~Benhamou, T.~Brathaug, T.~Christiansen, A.-L.
  Firle, A.~Fjeldly, L.~Frøyd, A.~Gleim, A.~Høiberget, C.~Meissner,
  G.~Nygård, J.~Olsen, H.~Paulshus, T.~Rasmussen, E.~Rishoff, J.~O. Skogås,
  \href{https://arxiv.org/abs/2304.11405}{Digital {Twins} in {Wind} {Energy}:
  {Emerging} {Technologies} and {Industry}-{Informed} {Future} {Directions}}
  (2023), Publisher: arXiv Version Number: 1.
\newblock

\bibitem{Pawar2021haa}
S.~Pawar, S.~E. Ahmed, O.~San, A.~Rasheed,
  \href{https://iopscience.iop.org/article/10.1088/1742-6596/2018/1/012031}{Hybrid
  analysis and modeling for next generation of digital twins}, {\em Journal of
  Physics: Conference Series}, 2018:012031 (2021).
\newblock

\bibitem{San2021haa}
O.~San, A.~Rasheed, T.~Kvamsdal,
  \href{https://onlinelibrary.wiley.com/doi/10.1002/gamm.202100007}{Hybrid
  analysis and modeling, eclecticism, and multifidelity computing toward
  digital twin revolution}, {\em GAMM-Mitteilungen}, 44 (2021).
\newblock

\bibitem{Stadtmann2023doa}
F.~Stadtmann, H.~G. Wassertheurer, A.~Rasheed,
  \href{https://arxiv.org/abs/2304.01093}{Demonstration of a {Standalone},
  {Descriptive}, and {Predictive} {Digital} {Twin} of a {Floating} {Offshore}
  {Wind} {Turbine}} (2023), Publisher: arXiv Version Number: 1.
\newblock

\bibitem{hoydedata}
Kartverket, \href{https://www.hoydedata.no}{Høydedata}.

\bibitem{kartverket}
Kartverket, {Norge digitalt},
  \href{https://kartkatalog.geonorge.no/}{Geonorge}.

\bibitem{norgeibilder}
{Statens vegvesen}, {Norsk institutt for bioøkonomi}, {Statens kartverk},
  \href{https://norgeibilder.no}{Norge i bilder}.

\bibitem{metthredds}
{Norwegian Meteorological Institute},
  \href{https://thredds.met.no/thredds/catalog.html}{{MET} {Norway} {Thredds}
  {Service}}.

\bibitem{NorwegianMeteorologicalInstitutemd}
{Norwegian Meteorological Institute},
  \href{https://github.com/metno/NWPdocs/wiki/MEPS-dataset}{{MEPS}
  documentation}.

\bibitem{Rasheed2017wfm}
A.~Rasheed, M.~Tabib, J.~Kristiansen,
  \href{https://asmedigitalcollection.asme.org/OMAE/proceedings/OMAE2017/57786/Trondheim,%20Norway/282088}{Wind
  {Farm} {Modeling} in a {Realistic} {Environment} {Using} a {Multiscale}
  {Approach}}, in {\em Volume 10: {Ocean} {Renewable} {Energy}}, American
  Society of Mechanical Engineers, Trondheim, Norway, 2017, p. V010T09A051.
\newblock

\bibitem{Yakoub2023ieo}
G.~Yakoub, S.~Mathew, J.~Leal,
  \href{https://linkinghub.elsevier.com/retrieve/pii/S0360544222027797}{Intelligent
  estimation of wind farm performance with direct and indirect ‘point’
  forecasting approaches integrating several {NWP} models}, {\em Energy},
  263:125893 (2023).
\newblock

\bibitem{Bolton1980tco}
D.~Bolton,
  \href{http://journals.ametsoc.org/doi/10.1175/1520-0493(1980)108<1046:TCOEPT>2.0.CO;2}{The
  {Computation} of {Equivalent} {Potential} {Temperature}}, {\em Monthly
  Weather Review}, 108:1046--1053 (1980).
\newblock

\bibitem{ea}
{Energy Transition Alliance}, \href{https://www.entralliance.com/}{{ENTR}
  {Alliance}}.

\bibitem{2022ain}
APQP4Wind,
  \href{https://apqp4wind.org/news/apqp4wind-is-now-cooperating-with-timwind}{{APQP4Wind}
  is now cooperating with {TIM} {Wind}} (2022).

\bibitem{Tran2020ges}
D.~T. Tran, H.~Robinson, A.~Rasheed, O.~San, M.~Tabib, T.~Kvamsdal,
  \href{https://iopscience.iop.org/article/10.1088/1742-6596/1669/1/012029}{{GANs}
  enabled super-resolution reconstruction of wind field}, {\em Journal of
  Physics: Conference Series}, 1669:012029 (2020).
\newblock

\bibitem{Larsen2020ota}
T.~N. Larsen, \href{https://hdl.handle.net/11250/2780978}{On the applicability
  of a perceptually driven generative-adversarial framework for
  super-resolution of wind fields in complex terrain}, Ph.D. thesis, NTNU,
  Trondheim, Norway (2020).

\bibitem{Kolle2022tiw}
K.~Kölle, T.~Göçmen, P.~B. Garcia‐Rosa, V.~Petrović, I.~Eguinoa, T.~K.
  Vrana, Q.~Long, V.~Pettas, A.~Anand, T.~K. Barlas, N.~Cutululis, A.~Manjock,
  J.~O. Tande, R.~Ruisi, E.~Bossanyi,
  \href{https://onlinelibrary.wiley.com/doi/10.1002/adc2.105}{Towards
  integrated wind farm control: {Interfacing} farm flow and power plant
  controls}, {\em Advanced Control for Applications}, 4 (2022).
\newblock

\bibitem{FMENorthWindnrc}
{FME NorthWind}, \href{https://www.northwindresearch.no/}{Norwegian {Research}
  {Centre} on {Wind} {Energy}}.

\end{thebibliography}


\end{document}